# New insight into the diffusion of hydrogen and helium atoms in tungsten

Qing HOU[*], Ailin YANG, Mingjie QIU, Jiechao CUI, Lei ZHAI

*Key Laboratory for Radiation Physics and Technology (Ministry of Education), Institute of Nuclear Science and Technology, Sichuan University, Chengdu 610064, China*

***Corresponding author** email address: qhou@scu.edu.cn

## ABSTRACT

Based on the in-detail tracking of the movements of atoms in a large number of molecular dynamics simulation boxes, we find that the diffusion of H and He atoms in single-crystal W is composed of non-Markovian jumps. The waiting time distribution of the triggering of jumps is not the usually recognized exponential distribution, but a temperature-dependent power-law distribution. The power-law distribution of the waiting time may lead to clear ergodicity-breaking diffusion, a phenomenon that has formerly been reported to only occur in complex systems such as living biological cells or soft matter. The present finding provides an insightful new view for the analysis and simulation of the puzzling H/He behaviors in W that are observed in experiments. Our findings will inspire reconsiderations on how to bridge the multiscale theoretical predictions with experimental observations, not only for H/He in W but also for H/He in other materials.

**Keywords:** non-Markovian diffusion; ergodicity-breaking diffusion; molecular dynamics; multiscale simulation; hydrogen isotope in tungsten; helium in tungsten

## 1. Introduction

The diffusion of H and He atoms in solids is a fundamental process closely related to phenomena observed in many scientific and technology fields. The diffusion of H/He atoms in metals cannot be directly experimentally observed on the atomistic level. The commonly supposed atomistic picture for the diffusion of H/He atoms in materials is as follows: H/He atoms move following the Markovian chain of jumps between energetically stable or metastable states, and the waiting time of the triggering of jumps follows the exponential distribution $\varphi(t) = exp\,(-t/\tau)$, where the average jumping time $\tau$ can be deduced through harmonic transition state theory (HTST) [1,2]. This postulate suggests a "normal" diffusion mechanism that is adopted in multiple time–space scale models for simulations and the analysis of experimental observations [3-5].

We report here a novel picture via massive molecular dynamics (MD) simulations for H/He atoms diffusing in tungsten, in which the diffusion of H/He atoms comprises non-Markovian jumps with the waiting time of the triggering of jumps following a temperature-dependent power-law distribution, and the diffusion may exhibit non-ergodic features at low temperatures. Here, the diffusion of H/He in W is chosen for the focus of our investigation because W is the most promising plasma-facing material in nuclear fusion reactors [6] and is subjected to high-fluence bombardments of H/He isotope atoms [7]. The long-term behavior of H/He in W has important impacts on the secure performance of the plasma-facing components in nuclear fusion reactors [8-10]. However, contradictions exist even in the very fundamental experimental and theoretical results for H and He diffusion in W. For example, the usually accepted experimental activation energy of H diffusion in W is 0.39 eV [9,11] against activation energies of 0.2 [12,13], 0.21 [14], 0.33 [15], and 0.38 eV [16] obtained by first-principles calculations; the experimental activation energy of He diffusion in W is 0.28 eV [17] against the first-principles value of 0.06 eV [18]. This disagreement can be attributed to many factors either in experimental techniques or theoretical predictions. One important factor is the diffusion model that must be adopted for connecting the atomistic processes with experimental observables. Therefore, the results presented in this paper will inspire reconsiderations on how to bridge the multiscale theoretical predictions with experimental observations, not only for H/He in W but also for H/He in other metals.

## 2. Method

Our MD simulation study on the diffusion process of H/He in W was conducted using our GPU-based molecular dynamics package, which can efficiently simulate the evolution of many simulation boxes in parallel [19]. Unless specifically stated, the key simulation setups applied for either H in W or He in W are outlined as follows. The adopted potentials were the embedded-atom-method potential that was developed and denoted as EAM1 by Bonny *et al.* [20] for the W–H–He system on the basis of the W–W potential of Marinica *et al.* [21]. The cutoff distance of the potentials was chosen to be identical to that used in making the potentials. Thousands cubic simulation boxes with side length $10a_0$ were initially constructed with 2000 W atoms on the bcc lattices and a H/He atom on each tetrahedron site, where $a_0$ is the lattice constant adopted by Marinica *et al.* for obtaining the W–W potential [21]. We thermalized the boxes to a preset temperature by repeating the thermalization–relaxation cycle 100 times. In each cycle, we generated the atomic velocities by sampling according to the Maxwell distribution and let the boxes relax for 0.1 ps. After the boxes reached thermal equilibrium at the preset temperature, the boxes were allowed to freely evolve, and only the atomic trajectories obtained in this stage would be used for analysis thereafter. Considering that the H/He atom is light and its frequency of thermal fluctuation is high, the timestep was set to 0.1 fs. The evolution time and the frequency of recording the instant status of the simulation boxes were set according to the quantities to be analyzed, as given in the results section. Because the periodic boundary conditions were always applied in all three directions, the atomic displacements, which could exceed the box boundary, were independently recorded in addition to the atomic positions constrained in the boxes.

In addition to recording the atomic displacements to be used for calculating the diffusion coefficients of H/He atoms, we also explored the waiting times of the triggering of jumps of H/He atoms. To identify the jump events more accurately, replicas of the simulation boxes were generated and were quenched to zero temperature by the steepest-descent algorithm. A jump event is counted if a tetrahedron site that the H/He atom is in changes, and the waiting time is defined by the time difference between two successive jump events.

## 3. Results

Figure 1 shows the representative trajectories of H and He atoms in W at temperatures of 400 and 600 K. The trajectories were specially generated by using a very fine time step (10 time steps, corresponding to 1 fs) to record the positions of H/He atoms. From a cursory inspection of this figure, we can see that the H/He atoms are bound in the tetrahedron site for most of the time and can climb over the barrier between two adjacent tetrahedron sites in a short time less than 100 fs; this is not unusual. However, we can also see that the paths of the H/He atoms moving through the barrier may not be the adiabatic path that is suggested by HTST and can be obtained by static mechanics (SM) using certain algorithms, for example, the nudged-elastic-band (NEB ) method [22,23], by searching saddle points on the potential surface. Most of the paths observed here are twisted, in contrast to the adiabatic path that is predicted by SM to be an almost visually straight connection between two adjacent tetrahedron sites. To make a more quantitative comparison between the adiabatic and non-adiabatic paths, we calculated the distances $R_{NN}$ between the H/He atoms and the nearest W atoms, as shown in Figure 2. The fluctuation and asymmetry features in a single jump that are observed for the $R_{NN}$ of MD but not $R_{NN}$ of SM indicate that the movements of the H/He atoms do not synchronize with the movements of the surrounding W atoms. The nonsynchronous movements of the H/He and W atoms can be understood because of the large mass difference between the H/He atom and W atom. It is also likely the cause of the large mass difference that, as observed in Figure 1, there are "hot" sites in which the kinetic energy of the H/He atoms on average is considerably high in comparison to the temperature. When a H/He atom is "hot," the H/He atom can conduct continuous jumps from tetrahedron site to tetrahedron site in a short time so that two consecutive jumps are correlated. These jumps, for example, indicated by arrows in Figure 1c) and d), appear as a ballistic characteristic. With the increase in temperature, such correlated jumps occur more frequently. At the same temperature, He atoms conduct ballistic-like jumps more often than H atoms do because the static diffusion barriers of H and He atoms in W obtained using the present potential are 0.21 and 0.09 eV, respectively [20].

The fast ballistic-like jumps look similar to the Levy flight [24]. The Levy flight is the mathematical model of random processes in which the distance diffusers move in a jump is described by a power-law distribution and could thus be a long distance. In contrast to Levy flight, however, the ballistic-like jumps here are truncated because of the collisions between the H/He atom and the surrounding host (W) atoms. Truncated Levy flights have been observed in simulation studies of Au nanoclusters diffusion on graphite surfaces [25,26] and adsorbed Ag atoms on Au surfaces [27]. Differing from the diffusion of adatoms on surfaces, H/He atoms here are in bulk W. In fact, we found by scrutinizing the trajectories that the jumping H/He atoms are more likely to bounce from the surrounding W atoms. To clarify this statement, we identified the jump events using the method described in Section 2. For both H and He, we conducted the identifying procedure every 5 fs in the total simulation time of 75 ps for 1000 or more simulation boxes. When the boxes are quenched to zero temperature, the jump direction and the deflection between two successive jumps can be well defined, as explained by the inset of Figure 3. There are four principal deflections denoted as T0-T1-T0, T0-T1-T2, T0-T1-T3, and T0-T1-T4. Among them, the T0-T1-T3 and T0-T1-T4 deflections are equivalent, where the T0-T1-T0 deflection refers to the H/He atoms bouncing back to the tetrahedron site where the jumps start from. Taking the count of the T0-T1-T0 deflection as a unit, the relative counts of deflections (RCD) are shown in Figure 3. If the jumps are Markovian, the RCD of T0-T1-T2 would be 1 and that of T0-T1-T3(T4) would be 2. However, Figure 3 reveals that the probabilities for T0-T1-T0, T0-T1-T2, T0-T1-T3, and T0-T1-T4 jump paths are not equal. The first preferred jump path is T0-T1-T0, namely, the H/He atoms have the largest probability of revisiting the site where they start the first jump. The secondary preferred jump path is T0-T1-T2. With increasing temperature, the RCD of T0-T1-T2 increases and the RCD of T0-T1-T3(T4) decreases. These results indicate further the ballistic characteristic of H/He movements. Moreover, the RCD would be atomic-mass-independent if the jumps were Markovian. Because of the ballistic characteristic, the RCD here appears to be atomic-mass-dependent. In Figure 3, the RCDs at 600 K obtained by artificially assigning the H/He atom with different atomic masses are also displayed. When we assign the H/He atom with a large atomic mass, the RCD of T0-T1-T2 and T0-T1-T3(T4) increases, indicating the suppression of the ballistic characteristic. Certainly, analyzing the correlations between the two successive jumps does not mean the chain of the correlated jumps contains only two jumps.

The interaction from the host atoms can also induce an effect on the other end of the scale, that is, the H/He atoms can be confined in a tetrahedron site for a long time depending on the temperature. Especially for H in W, the representative trajectories, which were selected for the visual inspection in Figure 1, are unlikely to be as representative. Among the simulation boxes, there are boxes in which the H atom remains in the same tetrahedron site without any jump into adjoining tetrahedron sites during the 80 ps. In Figures 4 and 5, we display the waiting time distribution (WTD) with the waiting time $t_w$ defined by the time interval between two identified successive jumps. This definition is justified, considering the time of the H/He climbing over the saddle point is short. The WTDs were normalized to the total counts of jump events identified at each temperature. From these figures, we can coarsely classify the jumps as fast if $t_w < 0.1$ ps and as slow if $t_w > 0.1$ ps. In the region of $t_w < 0.1$ ps, it is seen that the WTDs have one large peak around $t_w = 0.03$ ps for H atoms and around $t_w = 0.05$ ps for He atoms. The positions of the peaks are almost independent of the temperature. These peaks arise most likely from the ballistic-like jumps of the H/He atoms discussed above. Circumstantial evidence can be found in Figure S7 in which the WTDs of H in 600-K W obtained using different atomic masses for H atoms are compared. For tritium, the peak shifts to around $t_w = 0.045$ ps. When the H atoms are assigned the artificial atomic mass 16, the ballistic characteristic is suppressed and the peak diminishes.

If the fast ballistic-like jumps are the only sources for the diffusion of the H/He atoms, the diffusion would be normal only if the diffusion is measured on a timescale longer than the characteristic time of the truncated ballistic-like movements. However, the existence of the slow jumps may lead to different results.

In the region of $t_w > 0.1$ ps, the WTDs do not exponentially decrease but fit well to the power-law distribution:

$$F_p(t) = A/(\tau + t)^{1+\alpha} \quad (1),$$

which has a longer tail than the exponential distribution. The $\alpha$ value is temperature-dependent. The smaller the $\alpha$ value, the longer the tail. For both H and He atoms in W, the $\alpha$ value increases with the temperature. Theoretically [24,28-34], the $\alpha$ value of the power-law distribution has a critical value of 1. If $\alpha < 1$, the first moment of the distribution is divergent, suggesting no mean waiting time, as well as ergodicity-breaking and subdiffusion. As we will discuss, 2 is also a critical $\alpha$ value. From Figures 4 and 5, we observe that, for the He atoms, $\alpha$ larger than 2 is observed for the temperature $T \geq 600$ K. At $T = 400$ K, $\alpha$ is slightly less than 1. For the H atoms, $\alpha$ is smaller than 2 and larger than 1 at $1000$ K. With decreasing temperature, $\alpha$ is around 1 at $1000$ K and becomes much smaller than 1 at 600 and 400 K. These results signify that the diffusion of H/He atoms in W is likely ergodicity-breaking in a certain temperature range. We make further examinations of this by analyzing the mean squared displacement (MSD) of the H/He atom in W.

Figure 6 shows our calculation results from the MD data for the ensemble-averaged

MSD $\overline{\boldsymbol{r}^2(t)}$ defined as:

$$\overline{\boldsymbol{r}^2(t)} = \langle[\boldsymbol{r}(t) - \boldsymbol{r}(0)]^2\rangle \quad (2),$$

and the time-averaged MSD of single trajectories $\overline{\delta^2(\Delta; t_T)}$ defined as:

$$\overline{\delta^2\,(\Delta; t_T)} = \frac{1}{(t_T-\Delta)}\int_0^{t_T-\Delta}[\boldsymbol{r}(\Delta + t) - \boldsymbol{r}(t)]^2 dt \quad (3),$$

where $\Delta$ is the lag time and $t_T$ is the evolution time of the simulation boxes. If $\overline{\boldsymbol{r}^2(t)}$ and $\overline{\delta^2(\Delta; t_T)}$ (for $t_T \gg t$ and $\Delta = t$) are equivalent, the diffusion would be ergodic. Instead, the difference of $\overline{\delta^2(\Delta; t_T)}$ between different trajectories is an indicator of nonergodic diffusion [28,35]. We first investigate the MSDs for the He atoms in W. It is seen that, although the trajectory-to-trajectory deviation increases from high to low temperature, the $\overline{\delta^2(\Delta; t_T)}$ of He atoms in W converge well to $\overline{\boldsymbol{r}^2(t)}$, even at the temperature $T = 400$ K at which the $\alpha$ value of the WTD is 0.94. The large deviation between $\overline{\delta^2(\Delta; t_T)}$ of different trajectories at long lag time is understandable [35], because of the unavoidable statistical uncertainty when $\Delta$ is close to $t_T$ ($t_T$ is 300 ps for $T = 400$ K and 75 ps for $T \geq 600$ K).

We then investigate MSDs for the H atoms in W. The situation is very different. A strong temperature dependence of trajectory-to-trajectory scatter is observed for $\overline{\delta^2(\Delta; t_T)}$. The temperature is lower, and the trajectory-to-trajectory scatter of $\overline{\delta^2(\Delta; t_T)}$ is more pronounced, corresponding to the decreasing $\alpha$ value of the WTDs with decreasing temperature. For $T \geq 1000$ K, $\overline{\delta^2(\Delta; t_T)}$ tends to converge to $\overline{\boldsymbol{r}^2(t)}$ (also see Figure S8). At temperatures lower than 800 K, some H atoms remain quite localized in the whole evolution time, consistent with the observation when we visually inspect the trajectories. The large trajectory-to-trajectory scatter of $\overline{\delta^2(\Delta; t_T)}$ suggests H atoms conducting nonergodic diffusion in low-temperature W. A more quantitative analysis can be found in Figure S9, which displays the ergodicity-breaking parameter (EB) [28,36] that is defined as $EB(\Delta; t_T) = \langle\overline{\delta^2(\Delta; t_T)}^2\rangle / \langle\overline{\delta^2(\Delta; t_T)}\rangle^2 - 1$ and thought to be a sensitive measure of the deviation from ergodicity. The $EB(\Delta; t_T)$ provides quantified evidence for the diffusion of H in W at $T \leq 800$ K being ergodicity-breaking.

In theory, subdiffusion induced by ergodicity-breaking is indicated by the ensemble-averaged MSD having the form $\overline{\boldsymbol{r}^2(t)} \sim K_\alpha t^\alpha$ with $\alpha < 1$ [28]. However, the aforementioned fast ballistic-like jumps also contribute to the MSD. If these jumps are the only contributors, the ensemble-averaged MSDs $\overline{\boldsymbol{r}^2(t)}$ and ensemble-time-averaged MSDs $\langle\overline{\delta^2(\Delta; t_T)}\rangle$ would become normal when $t$ or $\Delta$ is longer than the characteristic time of the truncated ballistic-like movements. Figure 7 shows the logarithm of $\overline{\boldsymbol{r}^2(t)}$ and $\langle\overline{\delta^2(\Delta; t_T)}\rangle$ (correspondingly, the thick lines in Figure 6). It is seen that the $\overline{\boldsymbol{r}^2(t)}$ and $\langle\overline{\delta^2(\Delta; t_T)}\rangle$ of He in W are consistent, and both are linearly proportional to the time (lag time) even at $T = 400$ K, indicating the dominant contributions from the fast jumps.

In the case of H in W of $T \geq 600$ K, the $\overline{\boldsymbol{r}^2(t)}$ and $\langle\overline{\delta^2(\Delta; t_T)}\rangle$ agree well with each other and a good linear relationship is observed for both $\overline{\boldsymbol{r}^2(t)}$ vs. $t$ and $\langle\overline{\delta^2(\Delta; t_T)}\rangle$ vs. $\Delta$. Although the $EB(\Delta; t_T)$ clearly indicates that the diffusion of H in 600-K W is ergodicity-breaking, the ergodicity-breaking feature is covered by the contribution of truncated ballistic-like movements, and both the $\overline{\boldsymbol{r}^2(t)}$ and $\langle\overline{\delta^2(\Delta; t_T)}\rangle$ look "normal." At $T = 400$ K, both $\overline{\boldsymbol{r}^2(t)}$ and $\langle\overline{\delta^2(\Delta; t_T)}\rangle$ deviate from the linear relationship for $log(t) < 0.5$, namely, for $t$ (or $\Delta$) $< 3.16$ ps. One possible reason for this deviation arises from the fact that the MSD, a concept of continuous diffusion theory, is not applicable when there are few jumps. Above this time, a kink point around $log(t) = 1.5$ ($t = 32$ ps) is observed for $\overline{\boldsymbol{r}^2(t)}$. For below and above this point, $\overline{\boldsymbol{r}^2(t)}$ is $\sim t^{1.07}$ and $\sim t^{0.87}$, respectively, while $\langle\overline{\delta^2(\Delta; t_T)}\rangle \sim \Delta^{0.97}$, having a slight sublinear relationship.

The good linear relationship of $\langle\overline{\delta^2(\Delta; t_T)}\rangle$ vs. $\Delta$ (a property that may mislead one to normal diffusion), even in the case of ergodicity-breaking, is compatible with the inference of the continuous time random walk model [37-39], from which $\langle\overline{\delta^2(\Delta; t_T)}\rangle \sim \Delta / t_T^{1-\alpha}$ can be deduced [40,41]. The dependence of $\langle\overline{\delta^2(\Delta; t_T)}\rangle$ on $t_T$ provides a test of the so-called ageing of the diffusion process [28,36]. In Figure 8, we demonstrate the $\langle\overline{\delta^2(\Delta; t_T)}\rangle$ on $t_T$ for various chosen $\Delta$ for the H/He atoms in 400-K W. It can be seen that the diffusion of the H atoms in 400-K W exhibits the ageing feature. We also note that the $\alpha$ value deduced from the ageing dependence is 0.86–0.89, which is larger than the $\alpha$ value of 0.33 deduced from the WTD (see Figure 4), again, because of the mixed feature of the truncated ballistic-like movement and subdiffusion. For the He atoms in 400-K W, only very weakened ageing behavior can be observed.

## 4. Discussion

Anomalous diffusion is a phenomenon observed in various fields, as reviewed by Barkai *et al.* [33] and Metzler *et al.* [28]. Most of the observed anomalous diffusions take place in complex systems such as living biological cells or soft matter [42,43], and phenomenal kinetic models have been established to describe them [28,29]. Based on detailed analysis on a large number of atomic trajectories generated by molecular dynamics, we find for the first time that H/He atoms in single-crystal W may also exhibit diffusion behaviors away from our former knowledge. We can sketch the picture for the diffusion processes of the H/He atoms in W as follows. The H/He atoms at first remain at a bcc tetrahedron site for a waiting time until their jumps to the next tetrahedron sites are triggered. The triggered jumps could be ballistic-like and correlated likely from local and transient nonequilibrium while the whole system is in equilibrium, because of the large atomic mass ratio between the diffusors and host atoms. The waiting time of the triggering of jumps follows a temperature-dependent power-law distribution, which may lead to ergodicity-breaking and subdiffusive features, especially for the case of H atom diffusion in low-temperature W. (The subdiffusive feature of He atom diffusion in W is not as clear as that for H atoms. We have also performed some simulations not shown here, in which compressing pressure is applied to the simulation boxes. The primary results indicate clear ergodicity-breaking diffusion of He atoms in the compressed W.) This picture provides a new insightful view on some issues that have been confusing to us.

One of the issues is the formation of H isotope blisters experimentally observed in the plasma-facing W in nuclear fusion reactors [8]. The *ab initio* and MD calculations of Becquart *et al.* [44] and Henriksson *et al.* [45] show that the dilute interstitial H atoms cannot form clusters in W. Thus, the formation of H blisters is usually attributed to the preexisting vacancies or cavities in W where the H clusters can grow [8]. However, H blisters are also observed on the surfaces of single-crystal W irradiated by high-flux H atoms with low energies without directly producing vacancies [46-48]. Later, the MD and *ab initio* simulations of Liu *et al.* [49] and Hou *et al.* [50], in which very high H concentrations in W were artificially set, showed that the high concentration of H atoms may induce the formation of vacancies. However, we showed in a previous paper [51], in which the diffusion of H was taken as normal as usually thought, that the reachable concentration of H atoms would be a few orders of magnitude lower than that required to produce the vacancies event if the W is irradiated by the highest flux available in experiments. Now, this enigma can be clarified to some level by a simple analysis based on the finding of the present paper. Assuming the W surface is irradiated by H atoms with a constant flux $F_0$, the number of H atoms that remain at their original implantation positions without any jumps until time $t$ can be written as follows: $N_0(t) = F_0\left[t - \int_0^t dt_0 \int_{t_0}^t dt_1 W(t_1 - t_0)\right]$, where $W(t)$ is the WTD of jumps. If $W(t) \sim t$ is in the exponential relationship $W(t) = e^{-t/\tau}/\tau$, $N_0(t)$ would vanish for large $t$; if $W(t)$ takes the form of the power-law relation $W(t) = \alpha\tau^\alpha/(\tau + t)^{1+\alpha}$, $N_0(t) \approx F_0\,\tau(t/\tau)^{1-\alpha}/(1-\alpha)$ for $\alpha \neq 1$, or $N_0(t) \approx F_0 \tau ln\,(1 + t/\tau)$ for $\alpha = 1$ and $t \gg \tau$. This simple analysis sufficiently suggests that the H atoms would increasingly accumulate around their implantation positions with increasing irradiation time $t$ if $\alpha \leq 1$. In other words, blistering is likely unavoidable when W at low temperatures ($< 1000\ K$) is irradiated by the H isotope for a long time according to the WTDs obtained above. This prediction agrees qualitatively with the experimental observations. (In most of the experiments, D were used instead of H. Referring to Figure S7, the $\alpha$ value decreases with increasing atomic mass of H isotopes. This conclusion is true for D. We have calculated the WTDs also using other potentials in MD. The results shown in Figure S10 lead qualitatively to the same conclusion.)

The finding of the novel diffusion feature of the H/He atoms may have more significant impacts on methodological aspects. Because the timescale of dynamic processes that MD simulations are able to handle is usually less than a nanosecond, multiscale studies are indispensable. Among them, the kinetic Monte Carlo (KMC) and rate theory (RT)-based methods are two of the most applied approaches connecting what is extracted from MD simulations with the long-term consequences of the diffusion of impurities or defects in materials [13,52-55].

The KMC method is an event-driven method, based on the random walk model with the basic assumption that the jump events are Markovian. The correlated ballistic-like jumps appearing in the diffusion process of the H/He atoms clearly conflict with the basic assumption. The rate parameters adopted in the KMC are commonly constructed from *ab initio* or classical SM calculations in the framework of HTST [13,52].

The correlated ballistic-like jumps (referring to Figure 3) also conflict with the fundamental no-recrossing assumption of HTST [1].

Instead of event-driving in KMC, the evolution of systems in the RT-based methods is governed by master equations. Bedeaux *et al.* provided a mathematical analysis on the condition that should be satisfied if the solutions of the master equations are equivalent to solving the random walk problems [56]. Here, we can consider this an intuitive approach. In fact, the RT-based methods and the KMC methods can be respectively deemed as the simulations of two experimental approaches. In the first approach (simulated by RT methods), the events are detected and recorded in a fixed time interval $t_{\mathrm{I}}$. The number of recorded events, denoted $N_{\mathrm{I}}$, is variable in each run of the experiment because of the random feature of the events. Thus, the measured average time of an event occurring would be $t_{\mathrm{I}}/\langle N_{\mathrm{I}}\rangle$. In the second approach (simulated by KMC methods), the time interval $t_{\mathrm{II}}$, which is variable, is recorded for every fixed $N_{\mathrm{II}}$ event. The measured average time of an event occurring would be $\langle t_{\mathrm{II}}\rangle/N_{\mathrm{II}}$. Although a value for $\langle t_{\mathrm{II}}\rangle$ can always be calculated from the number of runs of the experiment, an accurate $\langle t_{\mathrm{II}}\rangle$ does not exist in the case that the time distribution of the event occurrence is $W(t) = \alpha\tau^{\alpha}/(\tau + t)^{1+\alpha}$ and $\alpha \leq 1$. In this case, the solution of the RT-based methods and the results of KMC are not comparable. In the case of $1 < \alpha < 2$, an accurate $\langle t_{\mathrm{II}}\rangle$ exists in principle. However, the second moment $\int_0^t W(t)t^2dt$ is $\sim t^{2-\alpha}$ for $t \gg \tau$, suggesting an increasing standard error with time. Thus, caution is recommended when the results of either RT-based methods and KMC methods are compared or the results of simulations and experiments are compared.

The focus of the present paper is for the diffusion of H/He in W. A comprehensive description of the evolution of H/He atoms in W needs also knowledge of other rate processes, for example the dissociation of H/He atoms from vacancies or clusters. The WTDs of these rate processes are also commonly assumed to be the exponential distribution. The findings of the present paper may inspire revisiting the WTDs of these rate processes. Moreover, the findings may also motivate similar investigations for H/He in other materials in which the long-term H/He behavior has important impacts on their macroscopic properties.

**Acknowledgements**: This work was supported partially by the National Natural Science Foundation of China (Grant No. 11775151).

## References

1 Truhlar, D. G., Garrett, B. C. & Klippenstein, S. J. Current Status of Transition-State Theory. *J. Phys. Chem.* **100**, 12771-12800 (1996).

2 Vineyard, G. H. Frequency factors and isotopic effects in solid state rate process. *J. Chem. Phys. Solids* **3**, 121-127 (1957).

3 Marian, J. *et al.* Recent advances in modeling and simulation of the exposure and response of tungsten to fusion energy conditions. *Nucl. Fusion* **57**, doi:10.1088/1741-4326/aa5e8d (2017).

4 Marian, J., Hoang, T., Fluss, M. & Hsiung, L. L. A review of helium–hydrogen synergistic effects in radiation damage observed in fusion energy steels and an interaction model to guide future understanding. *J. Nucl. Mater.* **462**, 409-421, doi:10.1016/j.jnucmat.2014.12.046 (2015).

5 Lu, G.-H., Zhou, H.-B. & Becquart, C. S. A review of modelling and simulation of hydrogen behaviour in tungsten at different scales. *Nucl. Fusion* **54**, doi:10.1088/0029-5515/54/8/086001 (2014).

6 Rieth, M. *et al.* Recent progress in research on tungsten materials for nuclear fusion applications in Europe. *J. Nucl. Mater.* **432**, 482-500, doi:10.1016/j.jnucmat.2012.08.018 (2013).

7 Greuner, H. *et al.* Investigation of European tungsten materials exposed to high heat flux H/He neutral beams. *J. Nucl. Mater.* **442**, S256-S260, doi:10.1016/j.jnucmat.2013.04.044 (2013).

8 Ueda, Y. *et al.* Baseline high heat flux and plasma facing materials for fusion. *Nucl. Fusion* **57**, doi:10.1088/1741-4326/aa6b60 (2017).

9 Tanabe, T. Review of hydrogen retention in tungsten. *Phys. Scr.* **T159**, doi:10.1088/0031-8949/2014/t159/014044 (2014).

10 Causey, R. A. & Venhaus, T. J. The use of tungsten in fusion reactors: a review of the hydrogen retention and migration properties. *Phys. Scr.* **T94**, 9-15 (2001).

11 Frauenfelder, R. Solution and Diffusion of Hydrogen in Tungsten. *J. Vac. Sci. Tech.* **6**, 388-397, doi:10.1116/1.1492699 (1969).

12 Liu, Y.-L., Zhang, Y., Luo, G. N. & Lu, G.-H. Structure, stability and diffusion of hydrogen in tungsten: A first-principles study. *J. Nucl. Mater.* **390-391**, 1032-1034, doi:10.1016/j.jnucmat.2009.01.277 (2009).

13 Fernandez, N., Ferro, Y. & Kato, D. Hydrogen diffusion and vacancies formation in tungsten: Density Functional Theory calculations and statistical models. *Acta Mater.* **94**, 307-318, doi:10.1016/j.actamat.2015.04.052 (2015).

14 Heinola, K. & Ahlgren, T. Diffusion of hydrogen in bcc tungsten studied with first principle calculations. *J. Appl. Phys.* **107**, doi:10.1063/1.3386515 (2010).

15 Xu, J. & Zhao, J. First-principles study of hydrogen in perfect tungsten crystal. *Nucl. Instr. Meth. Phys. Res. B* **267**, 3170-3174, doi:10.1016/j.nimb.2009.06.072 (2009).

16 Johnson, D. F. & Carter, E. A. Hydrogen in tungsten: Absorption, diffusion, vacancy trapping, and decohesion. *J. Mater. Res.* **25**, 315-327, doi:10.1557/jmr.2010.0036 (2011).

17 Amano, J. & Seidman, D. N. Diffusivity of 3He atoms in perfect tungsten crystals. *J. Appl.*

*Phys.* **56**, 983-992 (1984).

18 Becquart, C. S. & Domain, C. Migration energy of He in W revisited by ab initio calculations. *Phys. Rev. Lett.* **97**, 196402, doi:10.1103/PhysRevLett.97.196402 (2006).

19 Hou, Q. *et al.* Molecular dynamics simulations with many-body potentials on multiple GPUs—The implementation, package and performance. *Comput. Phys. Commun.* **184**, 2091-2101, doi:10.1016/j.cpc.2013.03.026 (2013).

20 Bonny, G., Grigorev, P. & Terentyev, D. On the binding of nanometric hydrogen-helium clusters in tungsten. *J. Phys. Condens Matter* **26**, 485001, doi:10.1088/0953-8984/26/48/485001 (2014).

21 Marinica, M. C. *et al.* Interatomic potentials for modelling radiation defects and dislocations in tungsten. *J. Phys. Condens Matter* **25**, 395502, doi:10.1088/0953-8984/25/39/395502 (2013).

22 Henkelman, r. & Jo´nsson, H. Improved tangent estimate in the nudged elastic band method for finding minimum energy paths and saddle points. *J. Chem. Phys.* **113**, 9978-9985 (2000).

23 Henkelman, G., Uberuaga, B. P. & Jo´nsson, H. A climbing image nudged elastic band method for finding saddle points and minimum energy paths. *J. Chem. Phys.* **113**, 9901-9904 (2011).

24 Zaburdaev, V., Denisov, S. & Klafter, J. Lévy walks. *Rev. Mod. Phys.* **87**, 483-530, doi:10.1103/RevModPhys.87.483 (2015).

25 Luedtke, W. D. & Landman, U. Slip diffusion and Levy flights of an adsorbed gold nanocluster. *Phys. Rev. Lett.* **82**, 3835-3838 (1999).

26 Maruyama, Y. Temperature dependence of Lévy-type stick-slip diffusion of a gold nanocluster on graphite. *Phys. Rev. B* **69**, doi:10.1103/PhysRevB.69.245408 (2004).

27 Boisvert, G. & Lewis, L. J. Self-diffusion on low-index metallic surfaces Ag and Au (100) and (111). *Phys. Rev. B* **54**, 2880-2889 (1996).

28 Metzler, R., Jeon, J. H., Cherstvy, A. G. & Barkai, E. Anomalous diffusion models and their properties: non-stationarity, non-ergodicity, and ageing at the centenary of single particle tracking. *Phys. Chem. Chem. Phys.* **16**, 24128-24164, doi:10.1039/c4cp03465a (2014).

29 Metzler, R. & Klafter, J. The random walks guid to anomaloues diffusion: a fractional dynamics approach. *Phys. Rep.* **339**, 1-77 (2000).

30 Jeon, J. H. & Metzler, R. Inequivalence of time and ensemble averages in ergodic systems: exponential versus power-law relaxation in confinement. *Phys. Rev. E* **85**, 021147, doi:10.1103/PhysRevE.85.021147 (2012).

31 Massignan, P. *et al.* Nonergodic subdiffusion from Brownian motion in an inhomogeneous medium. *Phys. Rev. Lett.* **112**, 150603, doi:10.1103/PhysRevLett.112.150603 (2014).

32 Cherstvy, A. G., Chechkin, A. V. & Metzler, R. Anomalous diffusion and ergodicity breaking in heterogeneous diffusion processes. *New J. Phys.* **15**, 083039 (2013).

33 Barkai, E., Garini, Y. & Metzler, R. Strange kinetics of single molecules in living cells. *Physics Today* **65**, 29-35, doi:10.1063/pt.3.1677 (2012).

34 Bel, G. & Barkai, E. Weak Ergodicity Breaking in the Continuous-Time Random Walk. *Phys. Rev. Lett.* **94**, doi:10.1103/PhysRevLett.94.240602 (2005).

35 Cherstvy, A. G. & Metzler, R. Nonergodicity, fluctuations, and criticality in heterogeneous diffusion processes. *Phys. Rev. E* **90**, 012134, doi:10.1103/PhysRevE.90.012134 (2014).

36 Cherstvy, A. G. & Metzler, R. Ergodicity breaking, ageing, and confinement in generalized diffusion processes with position and time dependent diffusivity. *J. Stat. Mech.* **2015**, doi:10.1088/1742-5468/2015/05/p05010 (2015).

37 Montroll, E. W. Random Walks on Lattices. III. Calculation of First-Passage Times with Application to Exciton Trapping on Photosynthetic Units. *J. Math. Phys.* **10**, 753-765, doi:10.1063/1.1664902 (1969).

38 Montroll, E. W. & Weiss, G. H. Random Walks on Lattices. II. *J. Math. Phys.* **6**, 167-181, doi:10.1063/1.1704269 (1965).

39 Montroll, E. W. & Scher, H. Random walks on lattices. IV. Continuous-time walks and influence of absorbing boundaries. *J. Stat. Mech.* **9**, 101-135 (1973).

40 Lubelski, A., Sokolov, I. M. & Klafter, J. Nonergodicity mimics inhomogeneity in single particle tracking. *Phys. Rev. Lett.* **100**, 250602, doi:10.1103/PhysRevLett.100.250602 (2008).

41 He, Y., Burov, S., Metzler, R. & Barkai, E. Random time-scale invariant diffusion and transport coefficients. *Phys. Rev. Lett.* **101**, 058101, doi:10.1103/PhysRevLett.101.058101 (2008).

42 Weigel, A. V., Simon, B., Tamkun, M. M. & Krapf, D. Ergodic and nonergodic processes coexist in the plasma membrane as observed by single-molecule tracking. *Proc. Natl. Acad. Sci.* **108**, 6438-6443, doi:10.1073/pnas.1016325108 (2011).

43 Jeon, J.-H., Leijnse, N., Oddershede, L. B. & Metzler, R. Anomalous diffusion and power-law relaxation of the time averaged mean squared displacement in worm-like micellar solutions. *New J. Phys.* **15**, 045011 (2013).

44 Becquart, C. S. & Domain, C. A density functional theory assessment of the clustering behaviour of He and H in tungsten. *J. Nucl. Mater.* **386-388**, 109-111, doi:10.1016/j.jnucmat.2008.12.085 (2009).

45 Henriksson, K. O. E., Nordlund, K., Krasheninnikov, A. & Keinonen, J. Difference in formation of hydrogen and helium clusters in tungsten. *Appl. Phys. Lett.* **87**, doi:10.1063/1.2103390 (2005).

46 Shu, W. M. *et al.* Microstructure dependence of deuterium retention and blistering in the near-surface region of tungsten exposed to high flux deuterium plasmas of 38 eV at 315 K. *Phys. Scr.* **T128**, 96-99, doi:10.1088/0031-8949/2007/t128/019 (2007).

47 Roszell, J. P., Davis, J. W. & Haasz, A. A. Temperature dependence of deuterium retention mechanisms in tungsten. *J. Nucl. Mater.* **429**, 48-54, doi:10.1016/j.jnucmat.2012.05.018 (2012).

48 Alimov, V. K. *et al.* Temperature dependence of surface morphology and deuterium retention in polycrystalline ITER-grade tungsten exposed to low-energy, high-flux D plasma. *Journal of Nuclear Materials* **420**, 519-524, doi:10.1016/j.jnucmat.2011.11.003 (2012).

49 Liu, Y.-N. *et al.* Mechanism of vacancy formation induced by hydrogen in tungsten. *AIP Advances* **3**, doi:10.1063/1.4849775 (2013).

50 Hou, J. *et al.* Hydrogen bubble nucleation by self-clustering: density functional theory and statistical model studies using tungsten as a model system. *Nucl. Fusion* **58**,

doi:10.1088/1741-4326/aacdb6 (2018).
51 Qiu, M. *et al.* Diffusion behavior of hydrogen isotopes in tungsten revisited by molecular dynamics simulations. *Chinese Phys. B* **27**, doi:10.1088/1674-1056/27/7/073103 (2018).
52 Xu, H., Osetsky, Y. N. & Stoller, R. E. Self-evolving atomistic kinetic Monte Carlo: fundamentals and applications. *J. Phys. Condens Matter* **24**, 375402, doi:10.1088/0953-8984/24/37/375402 (2012).
53 Becquart, C. S. & Domain, C. An object Kinetic Monte Carlo Simulation of the dynamics of helium and point defects in tungsten. *J. Nucl. Mater.* **385**, 223-227, doi:10.1016/j.jnucmat.2008.11.027 (2009).
54 Jourdan, T., Stoltz, G., Legoll, F. & Monasse, L. An accurate scheme to solve cluster dynamics equations using a Fokker–Planck approach. *Comput. Phys. Commun.* **207**, 170-178, doi:10.1016/j.cpc.2016.06.001 (2016).
55 Simmonds, M. J. *et al.* Expanding the capability of reaction-diffusion codes using pseudo traps and temperature partitioning: Applied to hydrogen uptake and release from tungsten. *Journal of Nuclear Materials* **508**, 472-480, doi:10.1016/j.jnucmat.2018.05.080 (2018).
56 Bedeaux, D., Lakatos-Lindenberg, K. & Shuler, K. E. On the Relation between Master Equations and Random Walks and Their Solutions. *J. Math. Phys.* **12**, 2116-2123, doi:10.1063/1.1665510 (1971).

**Figure 1**. x-View of representative trajectories of H and He atoms in W at temperatures of 400 and 600 K in a time period of 80 ps. The red spheres are W atoms at the bcc lattices, while the brown solid circles are tetrahedron sites. The small spheres are trajectories of the H/He atoms. Two adjacent points correspond to a time interval of 1 fs. The color of the small spheres is mapped to the kinetic energy from 0.03 to 0.4 eV. The corresponding y- and z-views of the trajectories are given in the Supplementary Materials (Fig. S1–S4). In a) and b), the arrows indicate the jumps for which the distances between H/He atoms and the eight nearest W atoms are calculated, and they are shown in Fig. 2. In c) and d), the arrows indicate example jumps that exhibit ballistic characteristics.

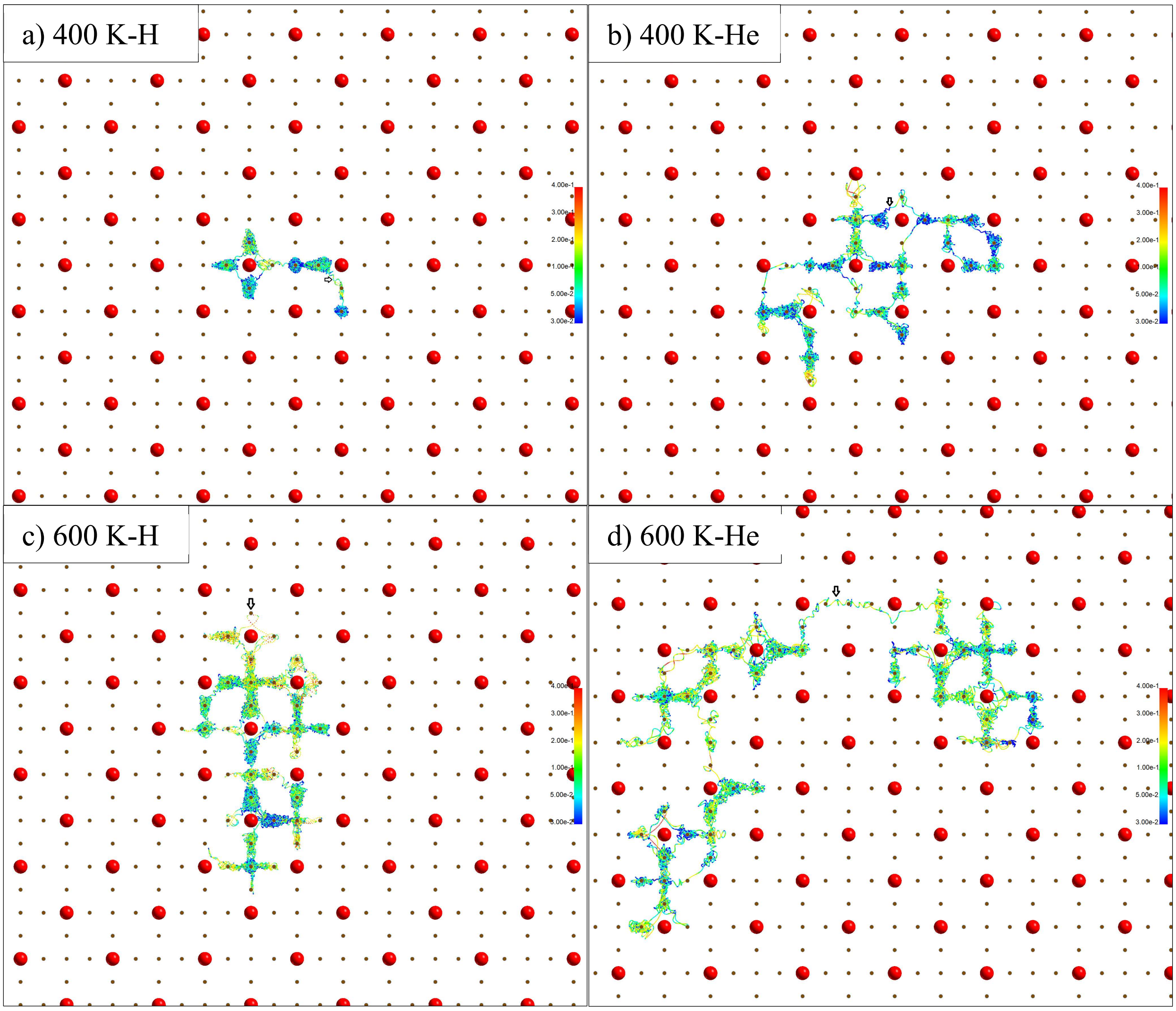

**Figure 2.** Comparison between the adiabatic and non-adiabatic diffusion path when a H/He atom jumps from a tetrahedron site to a neighboring tetrahedron site. The comparison is made by $R_{NN}$ and the distances between the H atom and its first eight nearest neighboring W atoms, as explained by the inset in Fig. 2a). The distances are sorted in the following incremental order: $R_{A1} < R_{A2} < R_{A3} < R_{A4} < R_{B1} < R_{B2} < R_{B3} < R_{B4}$. a) Adiabatic path obtained by NEB algorithm for the H atom jumping from the tetrahedron site defined by A1-A2 to that defined by A1-A2-A3-B1. Here, $R_{NN}$ is plotted as a function of NEB images. The intersection of $R_{A4}$ and $R_{B1}$ indicates a cross of the saddle point. For He, the path is similar; $R_{NN}$ as a function of time during the jump marked by the arrow in (b) Fig. 1a) and c) Fig. 1b). The $R_{NN}$ as a function of time in the whole 80 ps is given in the Supplementary Materials (Fig. S5 for H and Fig. S6 for He at 400 K).

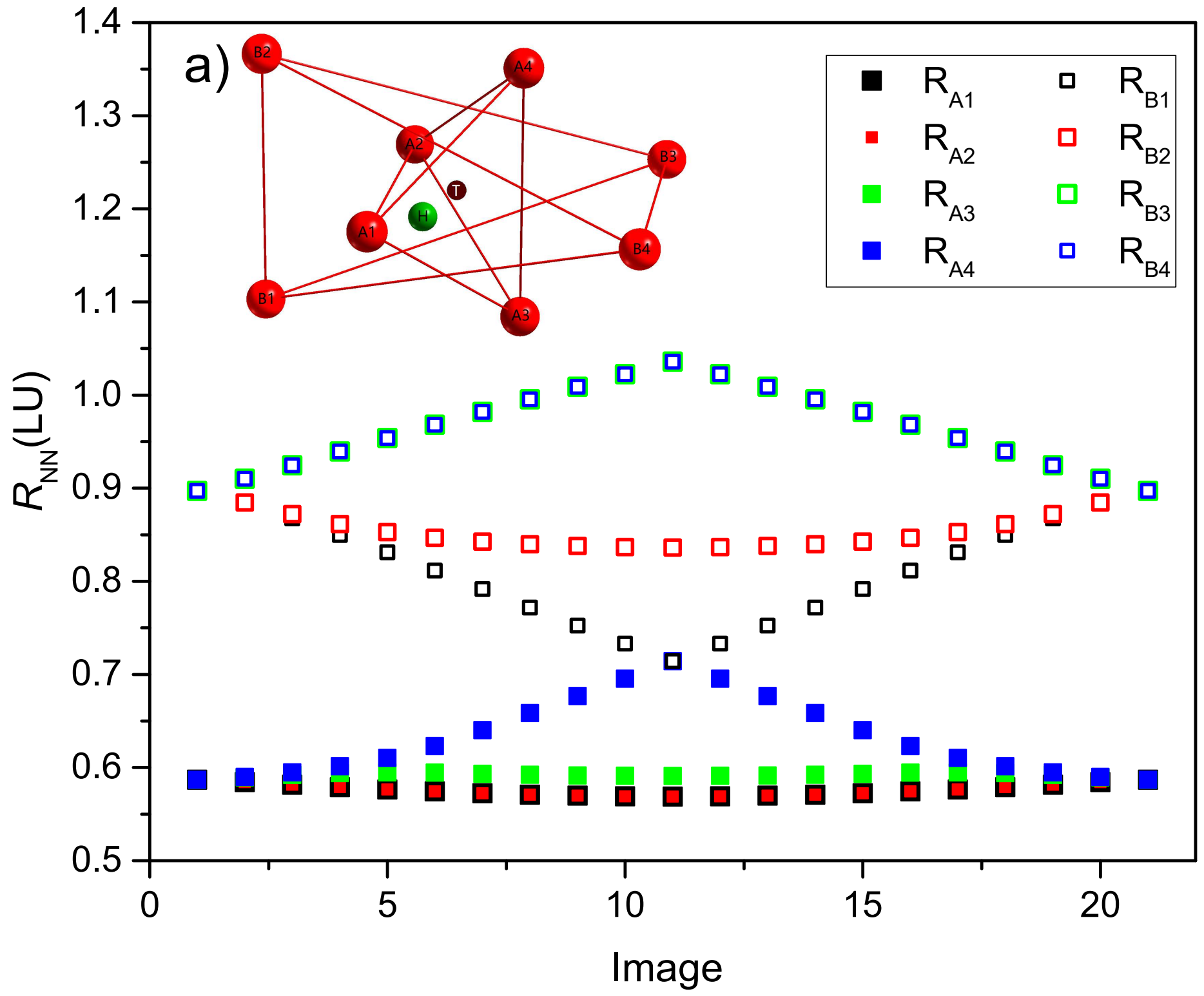

See next page

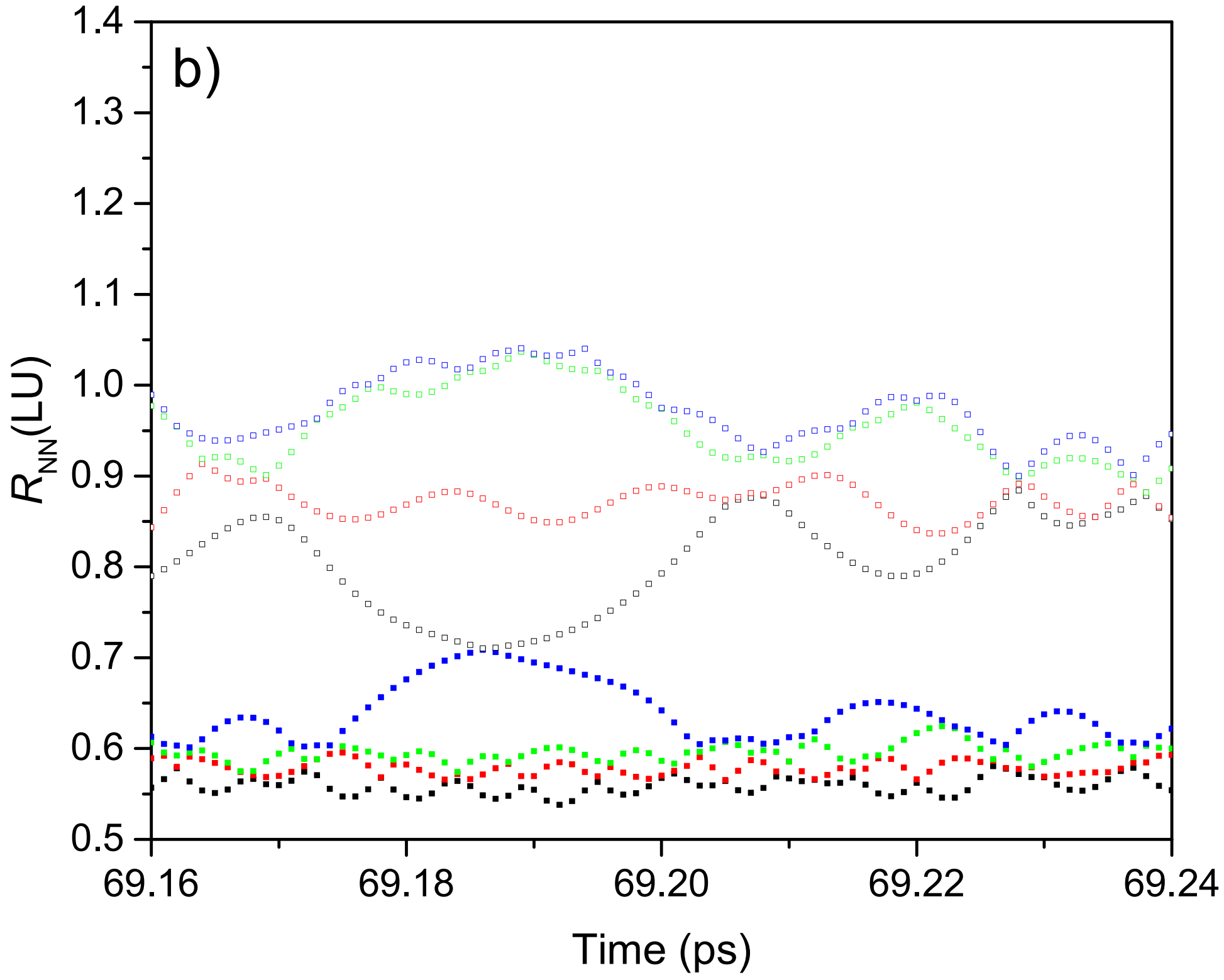

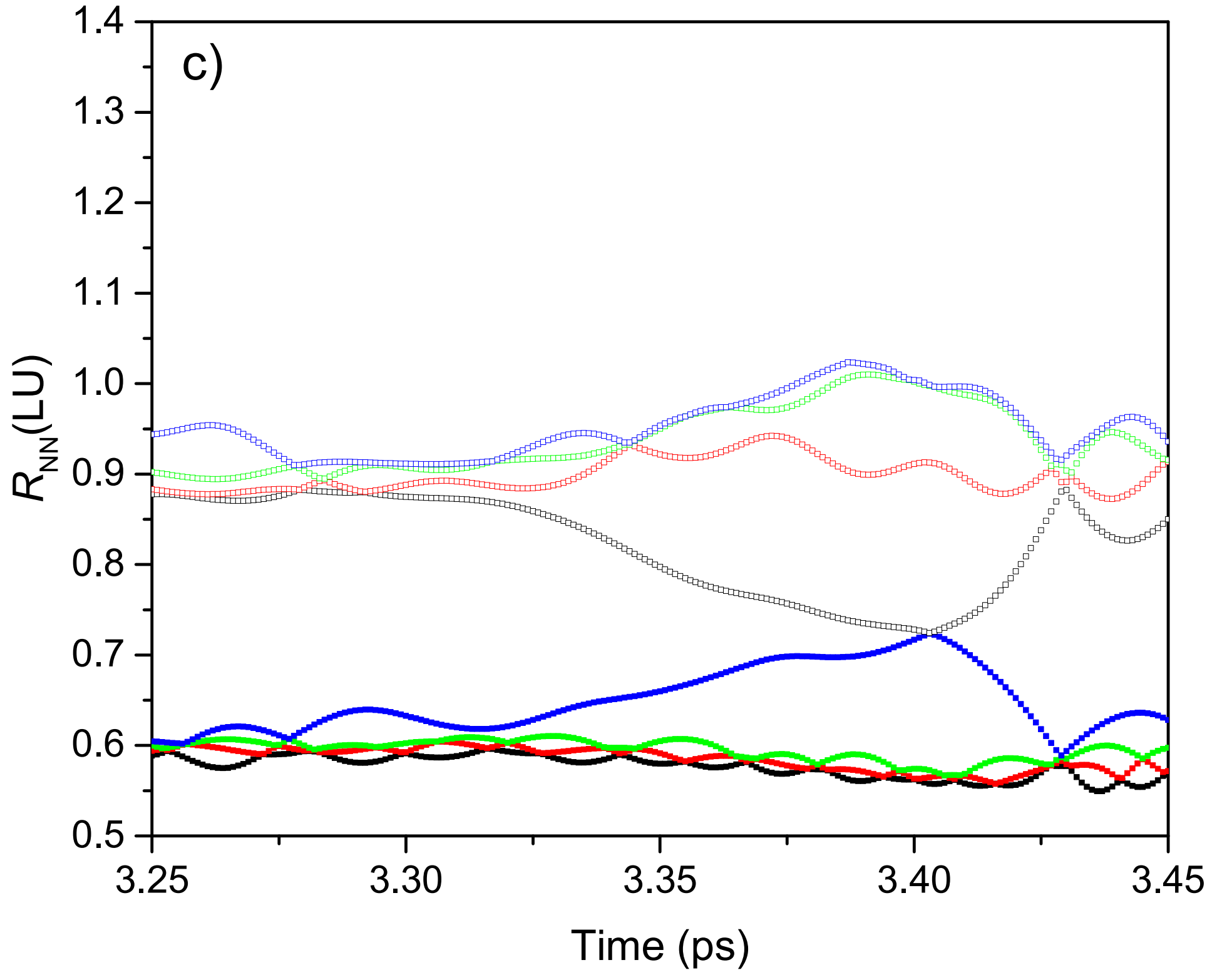

**Figure 3.** Relative count of the deflections between two successive jumps for a) H atom and b) He atom. The inset figure, in which red spheres are W atoms and golden spheres are tetrahedron sites, gives an explanation of the definition of the deflection. Assuming the first jump of the H/He atom is from the tetrahedron site T0 to T1, the jump vector is $V_{T0\text{-}T1}$. If the second jump, for example, is from T1 to T2, the deflection angle is the angle between $V_{T0\text{-}T1}$ and $V_{T1\text{-}T2}$. Corresponding to the four principal kinds of successive jumps T0-T1-T0, T0-T1-T2, and T0-T1-T3(T4), the cosine of the deflection is –1.0, 0.0, and 0.5 respectively, where T0-T1-T3(T4) denotes the summation of the equivalent successive jumps T0-T1-T3 and T0-T1-T4. The counts were normalized to the count of the T0-T1-T0 jump. The number of simulation boxes used to extract the data was 4000 at the temperature of 400 K and was 1000 at higher temperatures. The boxes evolved for 75 ps. At the given temperatures from 400 to 1200 K, the counts of the T0-T1-T0 jump were 6496, 12550, 38351, 73513, and 115472 for H, and they were 30927, 21189, 37224, 52988, and 66688 for He. Some results obtained using different atomic masses for the H/He atom are also shown to study the mass effect. M16 and M27 denote the artificial assignment of atomic masses 16 and 27 to the H/He atoms. Other potentials, denoted as EAM-Wang and FS, were also adopted to study the influence of the potentials.

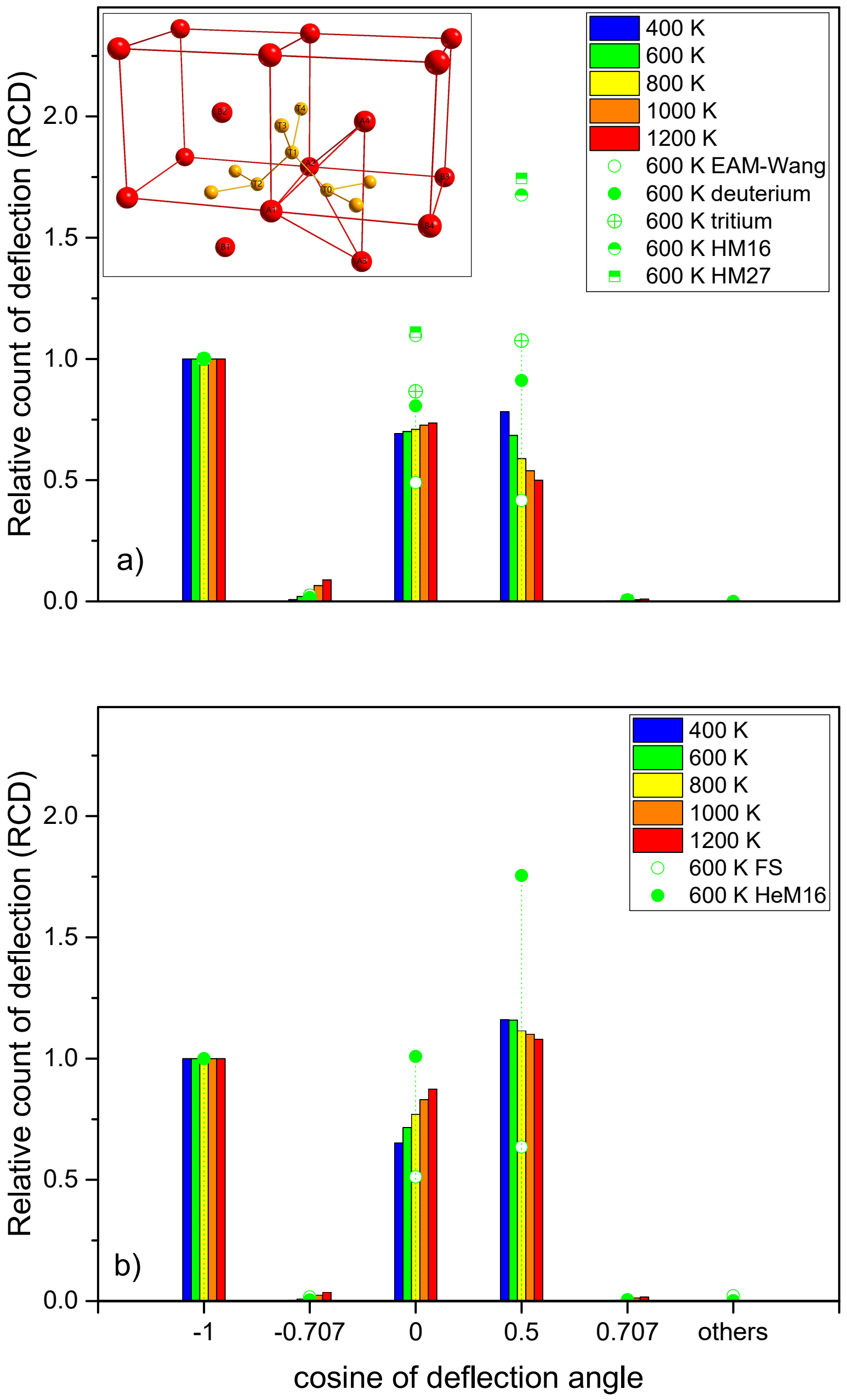

**Figure 4.** Distribution of waiting times for triggering a jump of an H atom at the following temperatures $T$: a) 400, b) 600, c) 800, and d) 1000 K. For $T$ = 400 K, 2000 simulation boxes were used to extract the waiting time. For $T$ above 400 K, 1000 simulation boxes were used. We conducted the procedure of identifying jump events every 5 fs (50 time steps). The distribution has been normalized to the count of jump events. The red solid curves indicate the power-law distribution, $F_p(t) = A/(\tau + t)^{1+\alpha}$, fit to the simulation data in range from 0.1 to 10 ps.

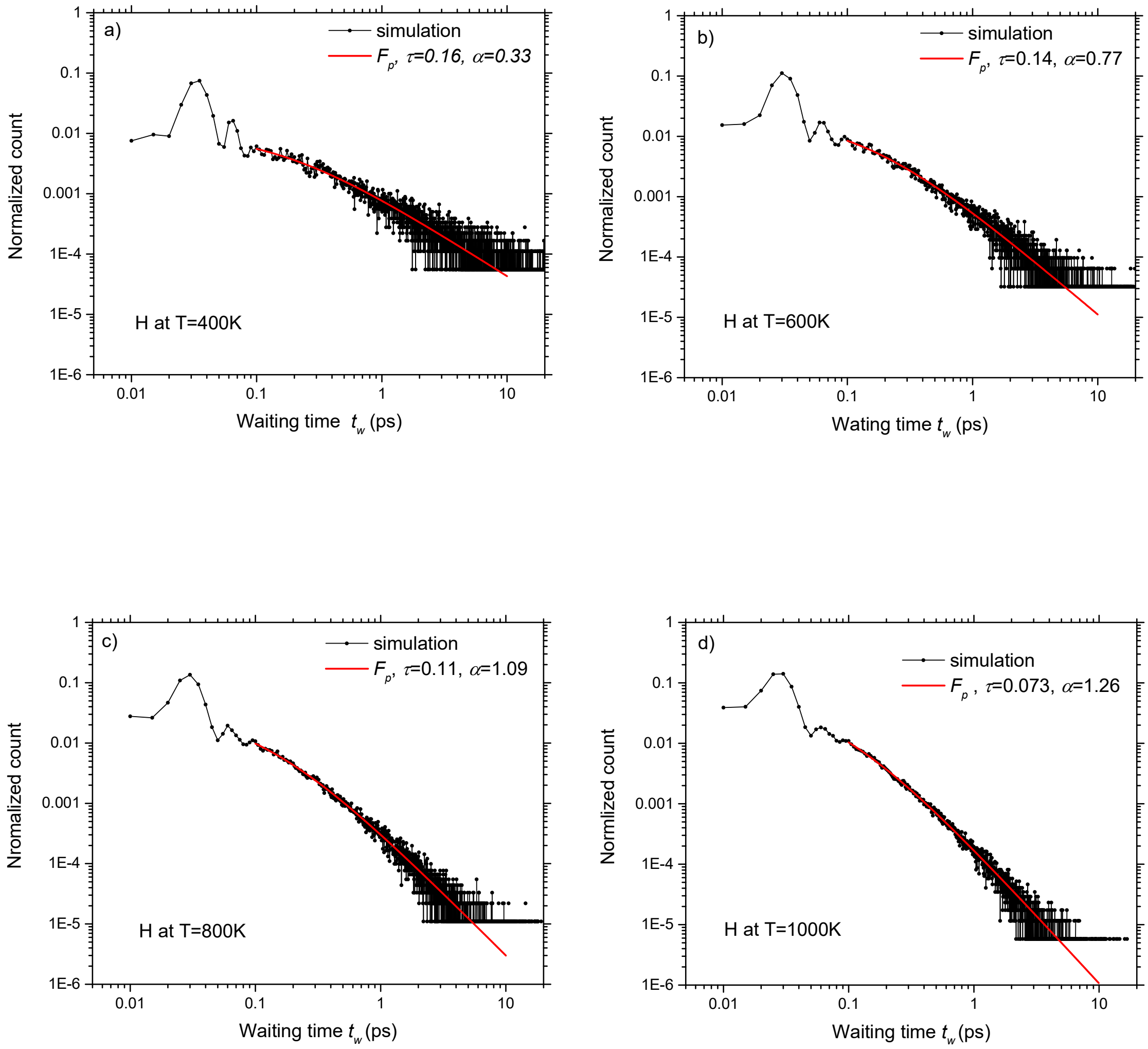

**Figure 5.** Distribution of waiting times for triggering a jump of an He atom at the following temperatures *T*: a) 400, b) 600, c) 800, and d) 1000 K. For *T* = 400 K, 2000 simulation boxes were used to extract the waiting time. For *T* above 400 K, 1000 simulation boxes were used. We conducted the procedure of identifying jump events every 5 fs (50 time steps). The distribution has been normalized to the count of jump events. The red solid curves indicate the power-law distribution, $F_p(t) = A/(\tau + t)^{1+\alpha}$, fit to the simulation data in the range from 0.1 to 10 ps. The green solid curve in d) is the exponential distribution, $F_e(t) = Aexp(-t/\tau)$, fit to the simulation data.

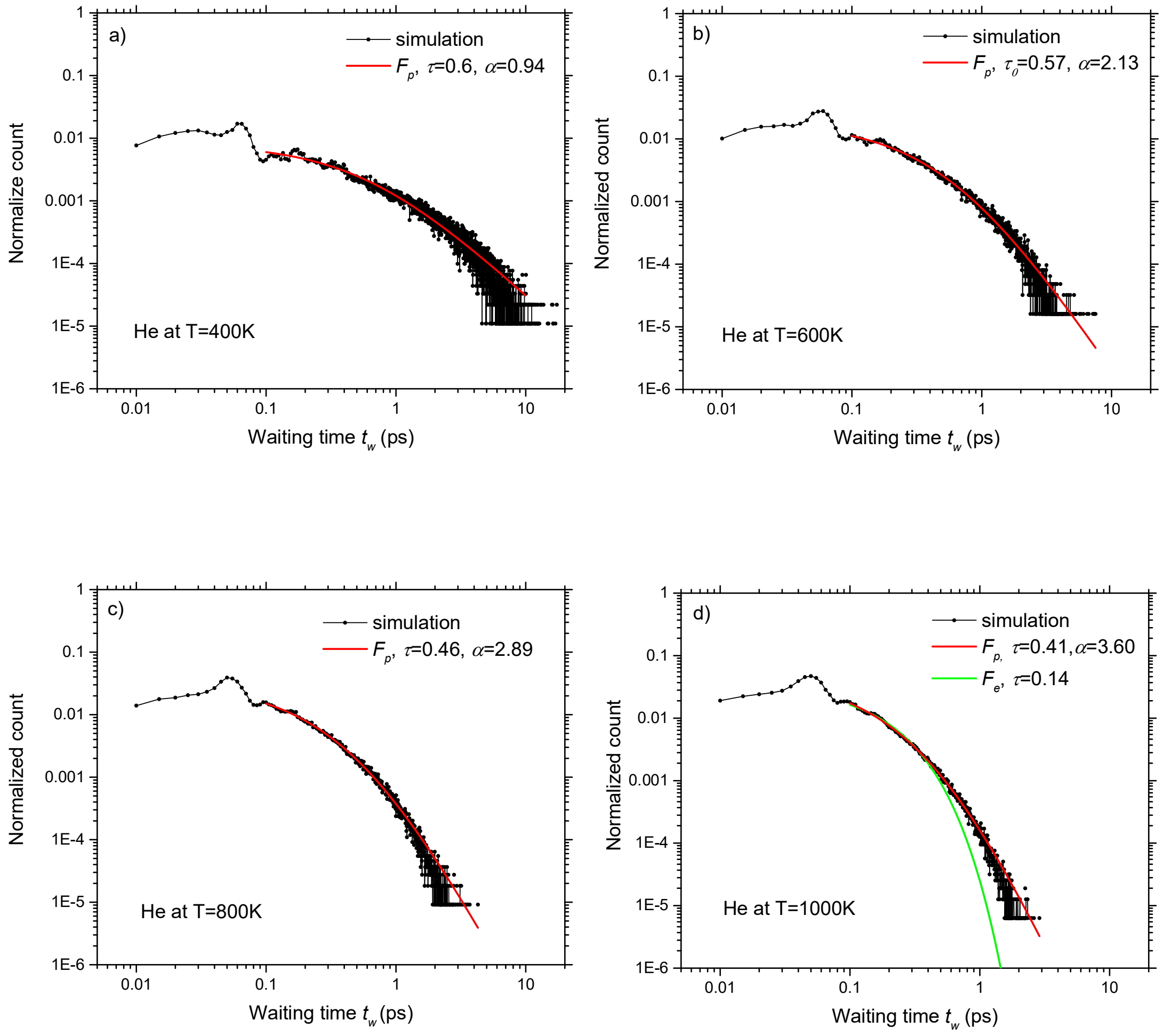

**Figure 6.** MSD vs. time for H (left column) and He (right column) in W at different temperatures. The unit of MSD is the square of the lattice length of W. The thin lines are the time-averaged MSDs of single trajectories defined by $\overline{\delta^2(\Delta; t_T)} = \int_0^{t_T-\Delta} [\boldsymbol{r}(\Delta + t) - \boldsymbol{r}(t)]^2 dt / (t_T - \Delta)$, where $\Delta$ and $t_T$ are respectively the lag time and the total evolution time of the simulation boxes. The thick green lines are $\langle \overline{\delta^2(\Delta; t_T)} \rangle$, the ensemble-average of $\overline{\delta^2(\Delta; t_T)}$. The thick dark lines are the ensemble-averaged MSD calculated by $\overline{\boldsymbol{r}^2(t)} = \langle [\boldsymbol{r}(t) - \boldsymbol{r}(0)]^2 \rangle$. For all temperatures, 4000 simulation boxes were used for MSD calculations, but displayed here are the $\overline{\delta^2(\Delta; t_T)}$ of the first 100 trajectories for each temperature. The configurations of the boxes were recorded every 0.5 ps. $t_T = 300$ ps at the temperature of 400 K and $t_T = 75$ ps at other temperatures. MSDs for higher temperatures can be found in Fig. S8.

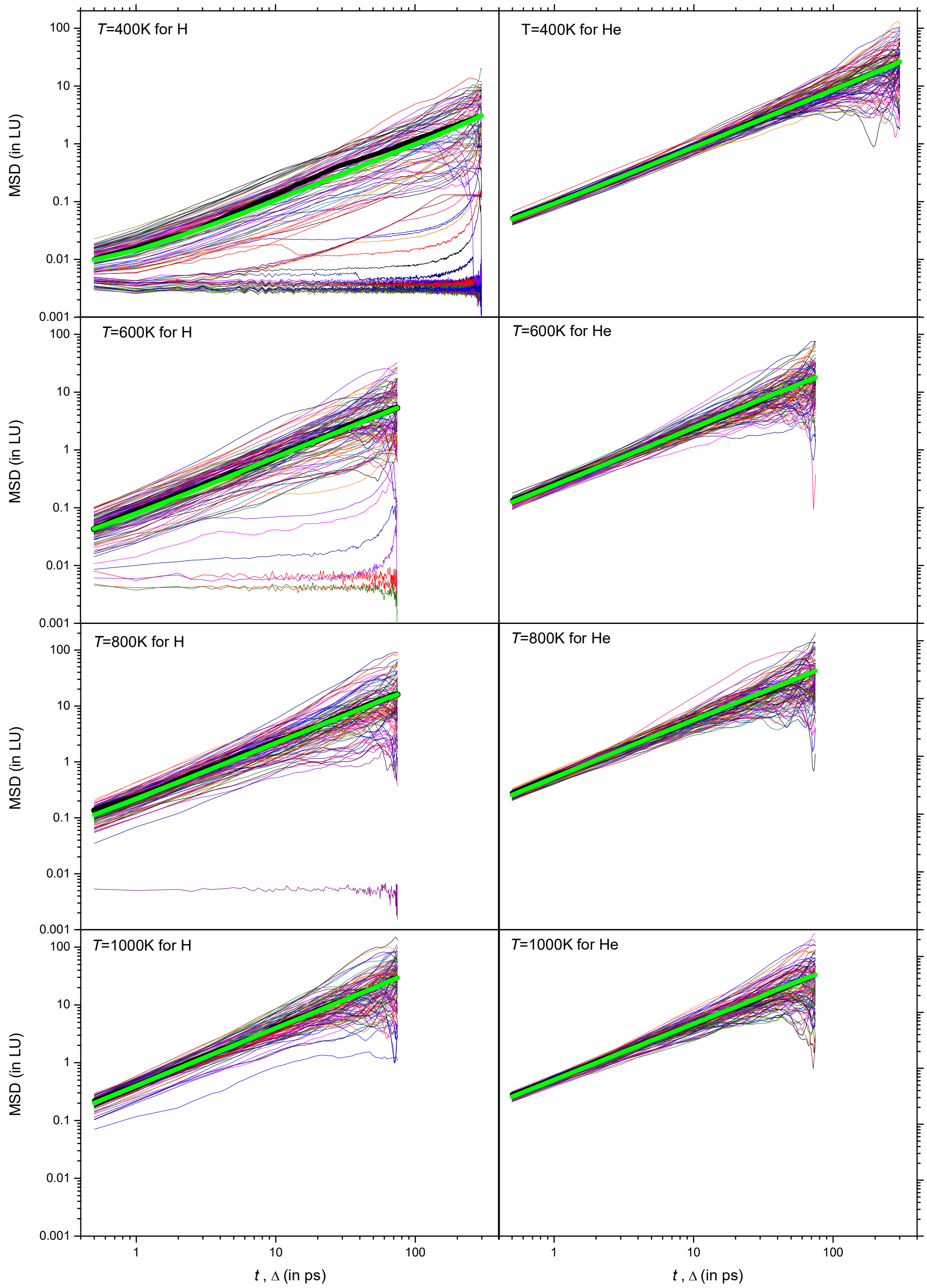

**Figure 7.** $log(\langle\overline{\delta^2\,(\Delta; t_T)}\rangle)$ vs. $log(\Delta)$ (solid symbols), and $log(\overline{\boldsymbol{r}^2(t)})$ vs log($t$) (hollow symbols). The solid lines are linear fits of $log(\langle\overline{\delta^2\,(\Delta; t_T)}\rangle)$, while the dashed lines are linear fits of $log(\overline{\boldsymbol{r}^2(t)})$ (only shown are for $T = 400$ K).

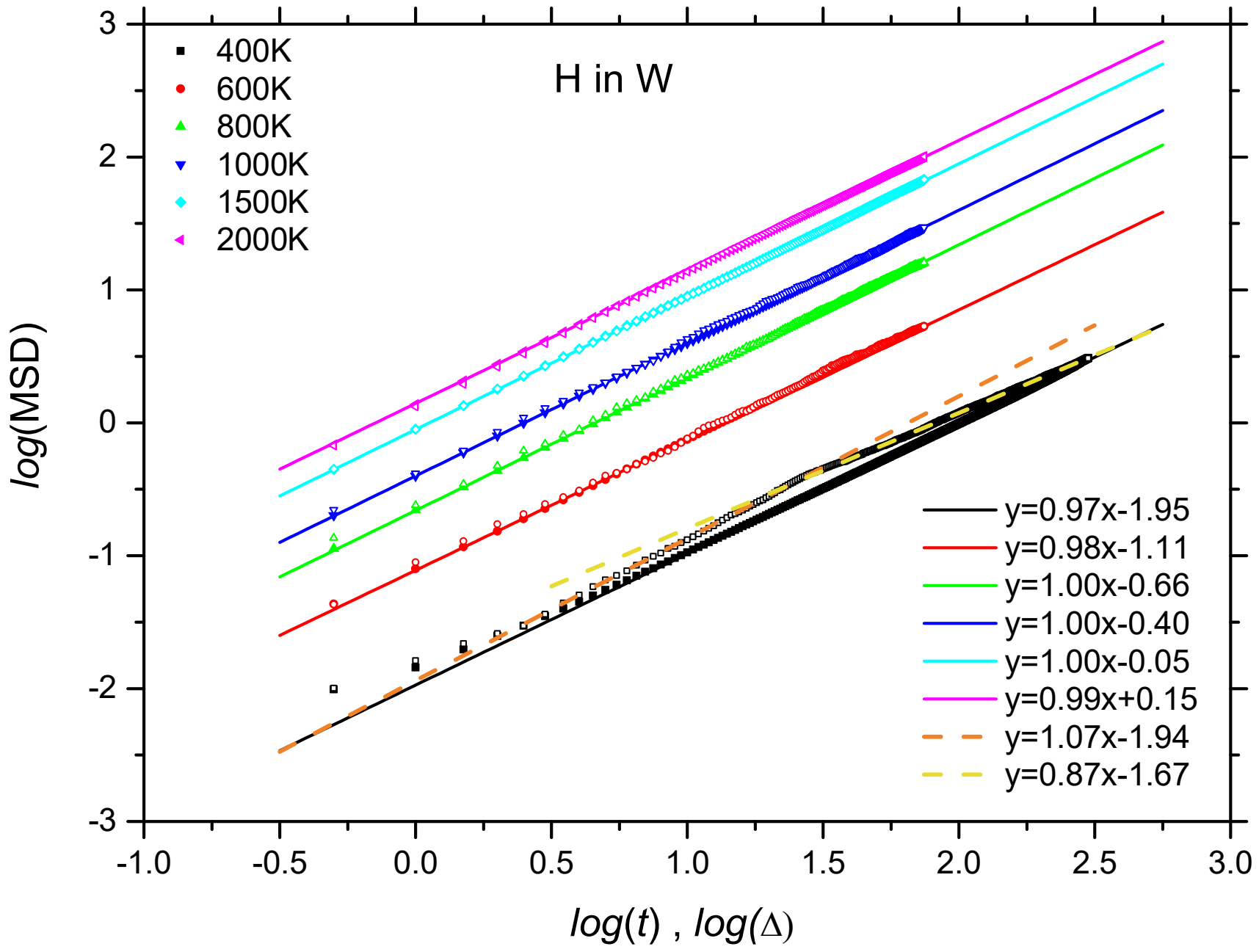

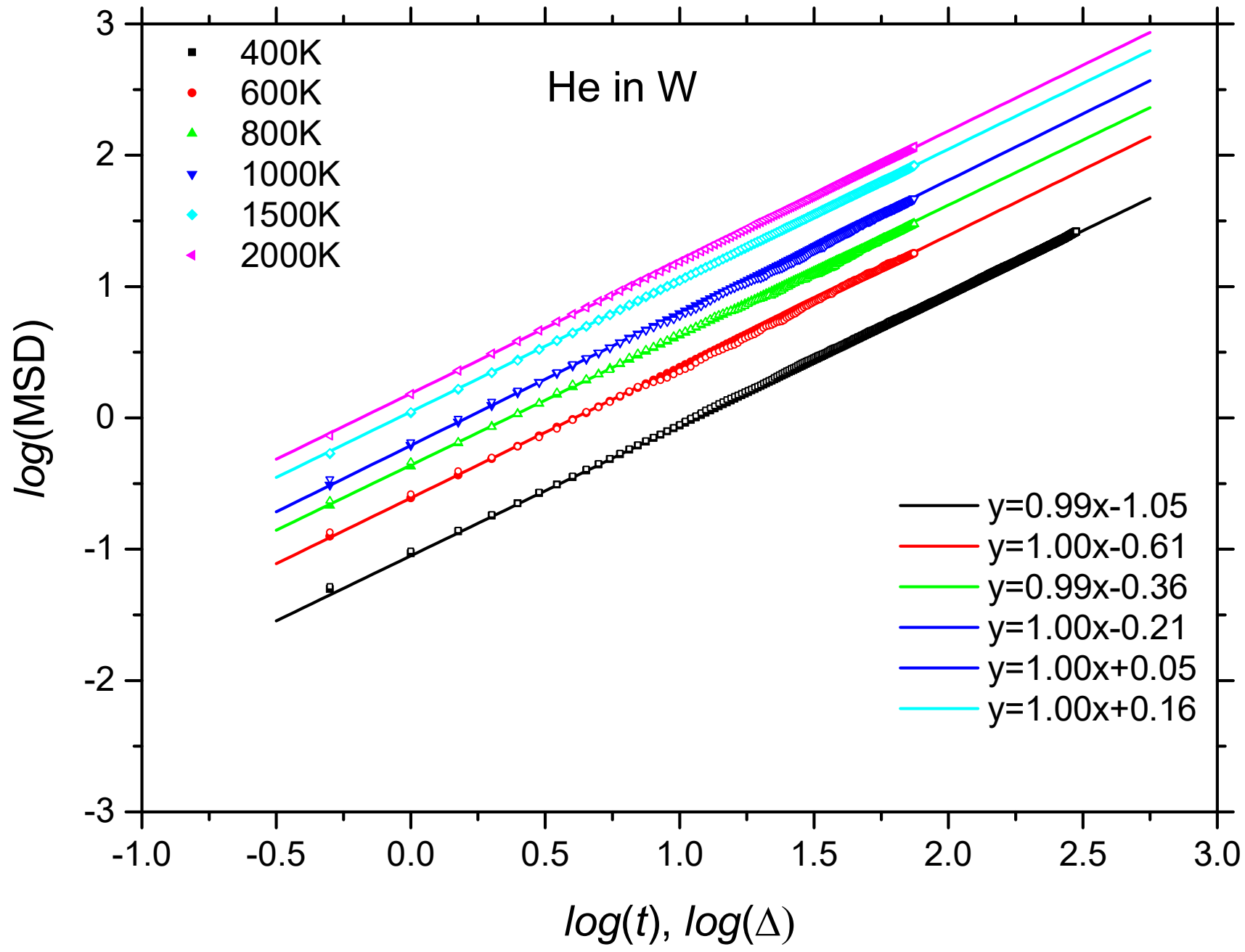

**Figure 8.** $\langle\overline{\delta^2(\Delta;t_T)}\rangle$ as a function of the evolution time $t_T$ for different lag times $\Delta$ (also see the caption of Fig. 6). From bottom to top, the corresponding lag time ranges from 2 to 100 ps in increments of 2 ps.

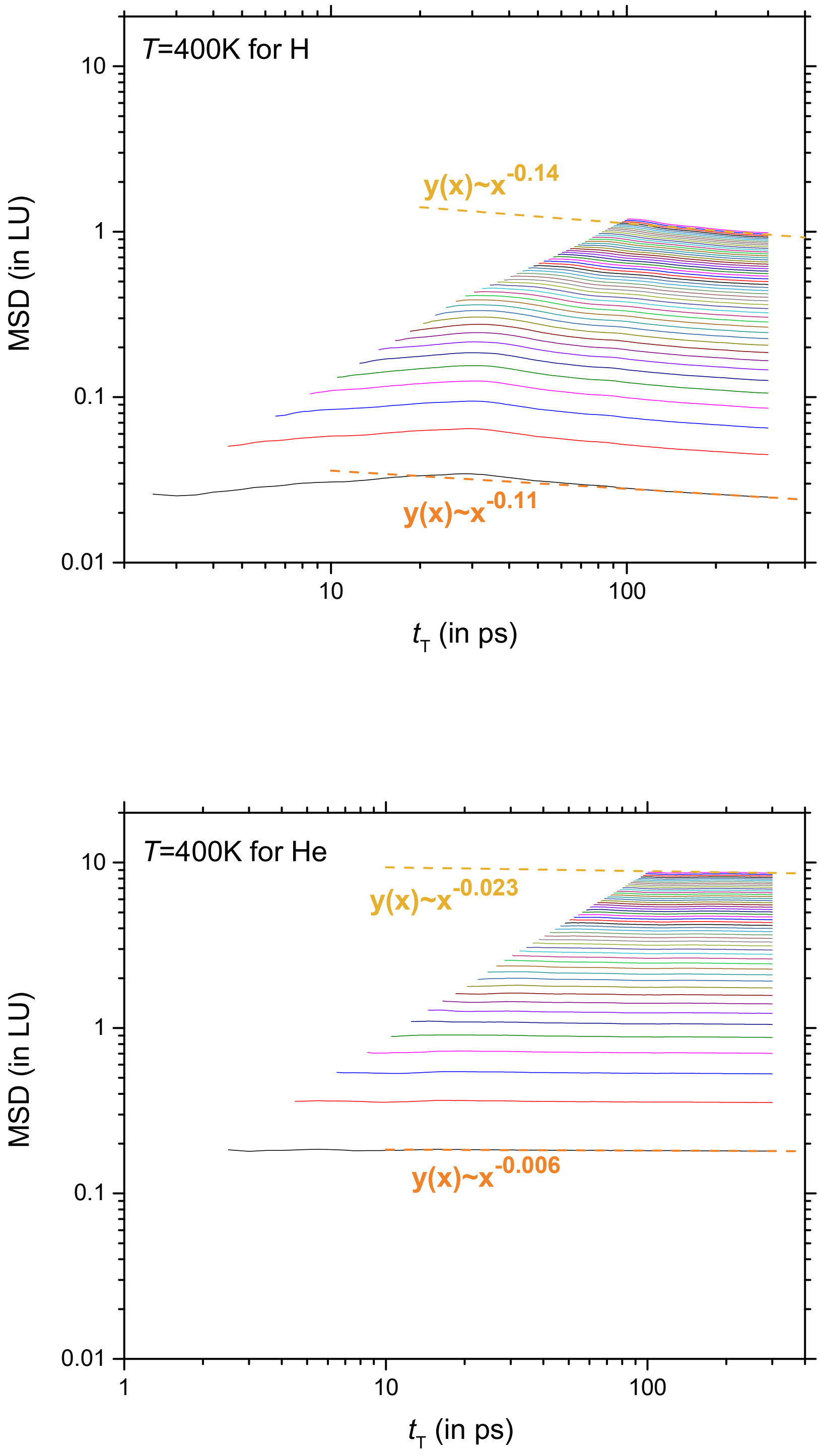